\documentclass[twocolumn]{aastex62}

\graphicspath{{./}{figures/}}

\received{August 2, 2018}
\revised{September 23, 2018}
\accepted{October 6, 2018}

\shorttitle{Space density of faint AGNs at $z\sim 4$}
\shortauthors{Boutsia et al.}

\begin{document}

\title{A high space density of L$^{*}$ Active Galactic Nuclei at $z\sim$4 in the COSMOS field}

\correspondingauthor{K. Boutsia}
\email{kboutsia@carnegiescience.edu}

\author[0000-0003-4432-5037]{K. Boutsia}
\affil{Carnegie Observatories, Las Campanas Observatory, Casilla 601, La Serena, Chile}
\author{A. Grazian}
\affiliation{INAF-Osservatorio Astronomico di Roma, via Frascati 33, I-00078, Monte Porzio Catone, Italy}
\author{E. Giallongo}
\affiliation{INAF-Osservatorio Astronomico di Roma, via Frascati 33, I-00078, Monte Porzio Catone, Italy}
\author{F. Fiore}
\affiliation{INAF-Osservatorio Astronomico di Trieste, via G.B. Tiepolo 11, I-34131, Trieste, Italy}
\author{F. Civano}
\affiliation{Harvard-Smithsonian Center for Astrophysics (CfA), 60 Garden Street, Cambridge, MA 02138, USA}

\begin{abstract}

Identifying the source population of ionizing radiation, responsible for the reionization 
of the universe, is currently a hotly debated subject with conflicting results. 
Studies of faint, high-redshift star-forming galaxies, in most cases, fail to detect 
enough escaping ionizing radiation to sustain the process. Recently, the capacity of 
bright quasi-stellar objects to ionize their surrounding medium has been confirmed 
also for faint active galactic nuclei (AGNs), which were found to display an escaping 
fraction of $\sim$74\% at $z\sim$4. Such levels of escaping radiation could 
sustain the required UV background, given the number density of faint AGNs is adequate. 
Thus, it is mandatory to accurately measure the luminosity function of faint AGNs 
(L$\sim$L$^{*}$) in the same redshift range. For this reason we have conducted a 
spectroscopic survey, using the wide field spectrograph IMACS at the 6.5m Baade 
Telescope, to determine the nature of our sample of faint AGN candidates in the 
COSMOS field. This sample was assembled using photometric redshifts, color, and X-ray 
information. We ended up with 16 spectroscopically confirmed AGNs at $3.6<z<4.2$ down 
to a magnitude of i$_{AB}$=23.0 for an area of 1.73 deg$^{2}$. This leads to an AGN 
space density of $\sim1.6\times10^{-6} Mpc^{-3}$ (corrected) at $z\sim$4 for an absolute 
magnitude of $M_{1450}=-23.5$. This is higher than previous measurements and seems to 
indicate that AGNs could make a substantial contribution to the ionizing background 
at $z\sim 4$. Assuming that AGN physical parameters remain unchanged at higher 
redshifts and fainter luminosities, these sources could be regarded as
the main drivers of cosmic reionization.

\end{abstract}


\section{Introduction} \label{sec:intro}
The reionization of the universe is the process where neutral hydrogen (H$_{I}$) becomes 
ionized and determines the transition from an opaque state to the transparent intergalactic 
medium (IGM) we observe today. Although the duration of this phase seems to be well 
established \citep{Fan06,Bec15,Planck}, the source population causing this effect 
is still elusive. The  debate is still ongoing as to whether the main contributors 
are faint, high-redshift, star-forming galaxies (SFGs) or active galactic nuclei (AGNs).
For both populations there are critical issues. The assumption for the galaxies being 
main contributors is that all faint SFGs should present an escape fraction of 
ionizing radiation {\it $f_{esc}$} of 10-20\% \citep{Fin15,Bow16}, which has not 
been observed so far, apart from in a handful of sources \citep{Sha16,Van16,Bian17,Ste18}.
Recent results \citep{Fle18,Jon18,Nai18,Tan18} do not provide a clear evidence, but 
indicate that it is difficult for the global SFG population to reach the 10-20\% level
of escape fraction required to drive the reionization process. The main objection against 
AGNs being the main contributors is that the number of bright quasars at $z>$4 is not 
enough \citep{Fan06,Cow09,HM12} and the number of faint AGNs at high redshifts is 
still not well constrained. Based on X-ray samples, at low luminosities in this redshift 
range, the space density of obscured AGNs is at least two times higher than the unobscured 
population \citep{Mar16b}, indicating that optically selected luminosity functions (LFs) 
could only be a lower limit.     

In order to answer the question of whether faint AGNs can contribute to the ionizing 
ultraviolet background (UVB), three aspects need to be quantified: (i) the exact level 
of the UVB; (ii) the fraction of ionizing radiation escaping these sources 
({\it $f_{esc}$}); and (iii) the faint slope of the AGN LF. 

Observations of the ionizing UVB intensity in the redshift range 2$<z<$5 and the global 
emissivity of ionizing photons indicate a relatively flat hydrogen photoionization 
rate \citep{BeB13}, but not spatially uniform \citep{Bos18}. Such large opacity 
fluctuations cannot be easily explained by low clustering populations like ultra-faint 
galaxies and could require the existence of rare bright sources at high redshift 
\citep{Bec15,Bec18,Cha15,Cha17}. As models start including larger AGN contributions, 
predicted temperatures are in agreement with observational constraints at $z\sim$4-6 
\citep{Kea18} (but see \citet{Puc18} for a different interpretation). 

Thus, a sizable population of faint (L$\sim$L$^{*}$) AGNs could contribute significantly 
to the UVB, as long as enough H$_{I}$ ionizing photons manage to escape the host galaxy 
\citep{MH15,Kha16}. In this respect, a recent study by \citet{Gra18}, using deep optical/UV 
spectroscopy, found that faint AGNs (L$\sim$L$^{*}$) at 3.6$<z<$4.2 present high escape 
fractions of ionizing radiation, with a mean value of 74\%. This is in agreement with 
similar studies of bright quasars (M$_{1450}\leq$ -26) at the same redshift range 
\citep{Cri16}, and there is no indication of dependence on absolute luminosities. This 
means that, if such results are extrapolated to higher redshift (5$<z<$7), the AGN 
contribution to the cosmic reionization process can become significant. At this point, 
knowledge of the exact number of faint AGNs at redshifts $z>4$ becomes crucial in 
accurately determining the level of this contribution.

Currently the consensus is that the LF for bright AGNs is well constrained, showing a 
peak at $z\sim$3 and then rapidly declining \citep{Bon07,Cro09}. However, for $z>$3 the 
debate is still open, with various studies presenting contradicting results. Works 
presented by \citet{Ike11} and \citet{Gli11} suggest that the number of faint AGNs 
at $z>$3 is higher than expected, producing a steeper slope at the faint end of the LF. 
But although the faint-end slopes are similar, the normalization factor $\Phi^{*}$ derived 
by \citet{Gli11} is three times higher than what calculated by \citet{Ike11} and 
subsequently reproduced by other studies \citep[i.e][]{Mas12,Aki18}. These latter 
studies report a strong decline in AGN numbers going from z=3 to z=4. In other words, 
there is still wide disagreement on the actual shape and normalization of the 
LF at $z\sim$4.

Work by \citet{Gia15}, including photometric and spectroscopic redshifts of X-ray-selected 
AGN candidates in the CANDELS GOODS-South region, has shown that at $z>$4 the probed AGN 
population could produce the necessary ionization rate to keep the IGM highly ionized 
\citep{MH15}. This result is still controversial, with recent works claiming the opposite 
\citep[i.e.][]{DAl17,Ric17,Aki18,Has18,Par18}. In fact, so far, the optical LFs at this 
redshift range and luminosities are based on a handful of spectroscopically confirmed 
sources (e.g., eight for \citet{Ike11} and five for \citet{Gia15}). Since the bulk of 
ionizing photons come from AGNs close to L$^{*}$, it is mandatory to measure their LF 
at $z>$4 in this luminosity range. For this reason, we started a pilot study in the 
COSMOS field, ideal for this kind of analysis thanks to its multi-wavelength catalog, 
X-ray, and radio coverage, which allows us to robustly select our AGN candidates. Here 
we present the bright part of our spectroscopically confirmed sample of 
intermediate-/low-luminosity AGNs, reaching an absolute magnitude of M$_{1450}$=-23 
and discuss a robust determination of the space density at $z\sim$4.

Throughout the paper we adopt the $\Lambda$ cold dark matter ($\Lambda$CMD) concordance 
cosmological model (H$_{0}$ = 70 km s$^{-1}$ Mpc$^{-1}$, $\Omega_{M}$ = 0.3, and 
$\Omega_{\Lambda}$ = 0.7). All magnitudes are in the AB system.

\section{AGN Candidate Selection} 
The selection of our sample is based on: (i) photometric redshifts, (ii) color-color 
selection, and (iii) X-ray emission. 

\begin{deluxetable*}{ccccccccc}
\tablecaption{Color-Color Candidates \label{tab:colorsel}}
\tablehead{
\colhead{ID} & \colhead{R.A.} &
\colhead{Decl.} & \colhead{i$_{AB}$} & \colhead{$z_{phot}$} &
\colhead{$z_{spec}$} & \colhead{(B$_{J}$-V$_{J}$)} &
\colhead{(r-i)} & \colhead{X-ray}\\}
\startdata
658294\tablenotemark{*} & 149.467350 &1.855592 &21.056&-1.000 & 4.174 &1.40&0.25 & no \\
1856470\tablenotemark{*}& 150.475680 &2.798362 &21.282&0.000 & 4.110 &1.42&0.32 & yes \\
1581239& 150.746170 &2.674495 &21.556&0.293 & -1.000&1.77&0.48 & no \\
507779& 150.485630 &1.871927 &22.034&0.605 & 4.450 &4.94&0.55 & yes \\
38736\tablenotemark{*}& 150.732540 &1.516127 &22.088&-1.000 & 4.183 &1.69&0.64 & no \\
1226535& 150.100980 &2.419435 &22.325&0.480 & 4.637 &1.68&0.43 & yes \\
422327 & 149.701500 &1.638375 &22.409&0.343 & 3.201 &1.54&0.14 & no \\
664641\tablenotemark{*} & 149.533720 &1.809260 &22.436&0.338 & 3.986 &1.69&0.30 & no \\
1163086\tablenotemark{*}& 150.703770 &2.370019 &22.444&-1.000 & 3.748 &1.44&0.25 & yes \\
330806\tablenotemark{*} & 150.107380 &1.759201 &22.555&3.848 & 4.140 &1.48&0.30 & yes \\
344777 & 150.188180 &1.664540 &22.634&0.392 & -1.000&1.89&-0.44 & no \\
1450499& 150.115830 &2.563627 &22.685&0.280 & 3.355 &1.94&0.63 & no \\
1687778& 150.006940 &2.779943 &22.715&0.437 & -1.000&1.96&0.44 & no \\
96886  & 150.289380 &1.559480 &22.765&3.860 & -1.000&1.77&0.27 & no \\
1573716& 150.729200 &2.739130 &22.783&0.376 & -1.000&1.35&0.48 & no \\
346317 & 150.205950 &1.654837 &22.800&0.352 & -1.000&1.450&-0.21 & no \\
1257518& 150.025190 &2.371214 &22.810&0.241 & -1.000&1.60&0.34 & no \\
1322738& 149.444050 &2.424602 &22.839&0.428 & -1.000&1.92&0.71 & no \\
1663056& 150.185000 &2.779340 &22.862&3.658 & -1.000&2.29&0.52 & no \\
1719143& 149.755390 &2.738555 &22.873&-1.000 & 3.535 &1.76&0.23 & yes \\
125420 & 150.222680 &1.510574 &22.898&0.181 & -1.000&1.83&0.53 & no \\
867305 & 149.446230 &2.115336 &22.950&0.651 & -1.000&2.11&0.71 & no \\
612661 & 149.838500 &1.829048 &23.011&4.229 & 4.351 &1.93&0.60  & no \\
\enddata
\tablenotetext{*}{Used for the LF}
\end{deluxetable*}

We use the photometric catalog and redshifts presented by \citet{Ilb09}. This is a 30-band 
catalog, spanning from NUV photometry to IRAC data, with calculated $z_{phot}$ in a 
region covering 1.73 deg$^{2}$ in COSMOS. The reported $z_{phot}$ dispersion 
is $\sigma_{(\Delta z)/(z_{s}+1)}$=0.007 at i$_{AB}<$22.5 and increases to 
$\sigma_{(\Delta z)/(z_{s}+1)}$=0.012 at i$_{AB}<$24. As discussed in \citet{Ilb09}, 
their $z_{phot}$ determination is mostly based on galaxy templates. However, as showed 
by \citet{Gia15}, for z$>$4 the accuracy on the photometric redshift estimate is weakly 
dependent on the adopted spectral libraries but it is mainly driven by the Lyman break 
feature at rest frame wavelength (912\AA). To take into account possible larger errors 
on photometric redshifts for the AGN population, we extended the redshift interval. Thus, 
we obtained a list of 42 candidates that have a photometric redshift estimate in the 
interval $3.0\leq z_{phot}\leq 5.0$ and a magnitude i$_{AB}<$23.0.

To increase our selection efficiency and mitigate shortcomings of the $z_{phot}$ technique, 
we include a color criterion. Since we have a wide number of bands available, initially 
we explored various combinations of color-color selections, i.e., (B$_{J}$-V$_{J}$) versus 
(r-i), (B$_{J}$-r) versus (r-i), (g-r) versus (r-i), (u$_{*}$-B$_{J}$) versus (r-i), and 
(u$_{*}$-g) versus (r-i). Cross-correlating those candidates with known AGNs from the 
literature, and after exploratory spectroscopy with 
LDSS-3\footnote{http://www.lco.cl/telescopes-information/magellan/instruments/ldss-3}, 
for this pilot study we narrowed down our selection to the most promising criterion, 
i.e. (B$_{J}$-V$_{J}$) versus (r-i). In the (B$_{J}$-V$_{J}$) versus (r-i) color-color 
diagram we consider as high-redshift AGN candidates the sources found in the locus 
delimited by: \\
\\
(B$_{J}$-V$_{J}$)$>$1.3 \\
and \\
(r-i)$\leq$0.60$\times$(B$_{J}$-V$_{J}$) - 0.30.\\
\\
With this criterion we obtained 23 candidates down to i$_{AB}$=23.0, summarized in 
Table \ref{tab:colorsel}. We decided not to put any constraints on the morphology, 
since the population of low-luminosity AGNs (M$_{1450}\sim$-23) includes Seyferts, 
where the host galaxy could be visible. 

\begin{figure}
\label{fig:spec}
\plotone{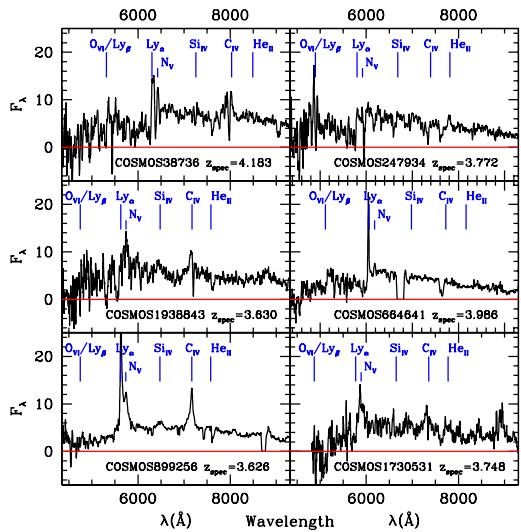}
\caption{The spectra of the six AGNs with $3.6\leq z\leq 4.2$ and i$_{AB}\leq $23.0 
discovered during our spectroscopic campaign with IMACS and LDSS-3. The red line corresponds 
to zero flux F$_{\lambda}$, in arbitrary units.} 
\end{figure}

There is a relatively small overlap between the candidates selected by the various methods. 
More specifically, only 7\% of the candidates selected by photometric redshifts are also 
included in the color-selected sample (three out of 42 objects), which is useful to 
increase our completeness. This is a clear advantage with respect to the works of 
\citet{Gli11} and \citet{Ike11} which only used four bands for their selections.

The final criterion for the creation of our sample was X-ray emission. In practice, we 
selected 38 sources detected in X-rays by deep Chandra observations in the COSMOS field 
\citep{Civ16} with $z_{phot}\geq$3 and a limiting magnitude i$_{AB}<$23. These photometric 
redshifts were provided by \citet{Mar16a} based on AGNs, galaxies or hybrid templates, 
as described in \citet{Sal11}. This sample consists both of type-1 and type-2 AGNs, and 
represents an unbiased census of the faint AGN population at this redshift. Only eight of 
the sources selected with the first two criteria present also emission in X-rays, while six 
candidates have been selected both by X-ray and color criteria. 

Our final sample consists of 92 AGN candidates with magnitudes i$_{AB}<$23, that have been 
selected by at least one of the methods mentioned above. Thanks to extensive spectroscopic 
campaigns carried out in the COSMOS field \citep[e.g.][]{Bru09,Ike11,Civ12,Mar16a,Hasin18}, 
22 of our 92 candidates have secure spectroscopic redshifts. To establish the nature of 
the remaining 70 sources (five of which have uncertain spectroscopic redshifts), we started 
an exploratory spectroscopic campaign at the Magellan Telescopes.  

\section{Spectroscopic Follow-up}  

We were awarded 2.5 nights with the wide-field Inamori-Magellan Areal Camera and 
Spectrograph \citep[IMACS,][]{Dre11} on the 6.5m Magellan-Baade telescope at Las 
Campanas Observatory to obtain spectra for our AGN candidates. We observed a total of five 
multi-slit masks with the IMACS f/2 camera (27$\arcmin$ diameter field of view) 
with total exposure times ranging from 3hr to 6hr, during dark time in 2018 February and 
March. The width of the slits was 1$\arcsec$.0 and the detector was used without binning 
(0$\arcsec$.2/pixel in the spatial direction). 

For the three 6hr masks we used the 300 line mm$^{-1}$ red-blazed grism (300\_26.7) with 
spectral sampling of 1.25{\AA} pixel$^{-1}$, while for the two 3hr masks we used the 200 
line mm$^{-1}$ grism that has a slightly lower resolution, sampling 2.04{\AA} pixel$^{-1}$. 
It is worth noting that the space density of our AGN candidates is such that only around 
three objects typically fall in an IMACS field of view at this magnitude limit.

We observed a total of 16 AGN candidates with magnitudes ranging from i$_{AB}=$20 to 23.0, 
and for 14 of them we obtained robust redshift determination at $z>3$, resulting in an 
efficiency of $\sim 88\%$, and two uncertain redshifts at $z>3$. Out of the sub-sample with 
secure redshifts, we found six AGNs in the redshift range $3.6\leq z_{spec}\leq 4.2$, and 
eight AGNs with a measured redshift of either $3.1<z_{spec}<3.6$ or $4.2<z_{spec}<4.7$.
The masks were completed with less reliable AGN candidates at $z\sim$4 obtained by relaxing 
the criteria mentioned above and exploring different color selections with respect to the 
(B$_{J}$-V$_{J}$) versus (r-i) one. We also added, as fillers, fainter candidates, 
i$_{AB}\leq$ 24.0, in order to explore the fainter regime of the AGN LF. 

After the spectroscopic campaign, 36 sources of the parent sample of 92 candidates had 
secure spectroscopic redshifts. In addition, seven sources had uncertain spectroscopic 
classification, either because of low signal-to-noise ratio or because only one line was 
visible. This left 49 candidates with no redshift information. Figure 1 shows the spectra 
of the six AGNs with $3.6\leq z_{spec}\leq 4.2$ and i$_{AB}\leq $23.0 discovered during 
our spectroscopic campaign with IMACS and LDSS-3. Most sources show characteristic AGN 
lines like Si$_{IV}$ and C$_{IV}$. The two sources that don not show Si$_{IV}$ and C$_{IV}$, 
have been included for other strong high-ionization lines that are also characteristic of 
AGNs. More specifically, COSMOS664641 has strong O$_{VI}$ and N$_{V}$ lines, while the 
C$_{IV}$ line falls in a region of telluric absorption and this could be the reason it is 
not observed. In the case of COSMOS247934 we also detect strong O$_{VI}$/Ly$\beta$, 
as well as He$_{II}$ lines, and it is relatively bright with $M_{1450}\sim -23.2$. The fact 
that these sources lack X-ray emission does not preclude their AGN nature, since there is 
a number of examples of AGNs not detected by deep X-ray surveys \citep{Ste02}. Moreover, 
their luminosities ($M_{1450}\le -23.5$) are another indication of their nuclear activity. 
In the subsequent analysis we only consider sources with secure spectroscopic redshifts 
either from our campaign or from the literature.

\begin{figure}
\label{fig:col}
\plotone{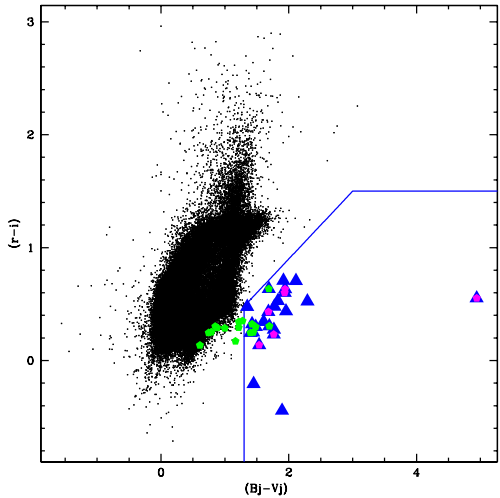}
\caption{Color-color diagram. Blue triangles show candidates selected by color. 
Pentagons show candidates selected by photometric redshifts or X-ray emission and 
confirmed by spectroscopy as AGNs. Green pentagons correspond to AGNs at 3.6$<z_{spec}<$4.2 
and magenta pentagons show confirmed AGNs that have either z$_{spec}<$3.6 or z$_{spec}>$4.2. 
Notice that half of the confirmed AGNs are found outside the color selection locus and that 
half of the color selected candidates still need to be observed.} 
\end{figure}

The distribution of our candidates in the color-color space can be seen in Figure 2. 
Here we plot the entire color-selected sample and indicate which sources were confirmed as 
AGNs, in the redshift range of interest, either after our spectroscopic campaign or from 
the literature. We also indicate sources that lie in the color locus but their spectroscopic 
redshifts are either $z_{spec}<$3.6 or $z_{spec}>$4.2. A detailed presentation of the 
full spectroscopic sample and a comprehensive description of the different color criteria 
are not the main aims of the present paper and will be discussed in a future work.

\begin{deluxetable*}{ccccccccc}
\tablecaption{Confirmed AGNs Used for Determining Space Density \label{tab:spaced}}
\tablehead{
\colhead{ID} & \colhead{R.A.} &
\colhead{Decl.} & \colhead{$i_{AB}$} & 
\colhead{$z_{spec}$} & \colhead{$r_{AB}$} &\colhead{M$_{1450}$} & References \\}
\startdata
38736 & 150.732540 & 1.516127 & 22.088 & 4.183 & 22.897 & -23.341   & our spectroscopy \\
247934 & 150.801300 & 1.657550 & 22.334 & 3.772 & 22.817 & -23.182  & our spectroscopy \\
330806 & 150.107380 & 1.759201 & 22.555 & 4.140 & 23.105 & -23.110  & \citet{Ike11} \\
658294 & 149.467350 & 1.855592 & 21.056 & 4.174 & 21.603 & -24.630  & \citet{Tru09} \\
664641 & 149.533720 & 1.809260 & 22.436 & 3.986 & 22.946 & -23.182  & our spectroscopy \\
899256 & 150.782210 & 2.285049 & 21.927 & 3.626 & 22.363 & -23.545  & our spectroscopy \\
1054048\tablenotemark{*} & 149.879200 & 2.225839 & 22.697 & 3.650 & 23.200 & -22.722  & \citet{Mar16a} \\
1159815 & 150.638440 & 2.391350 & 22.157 & 3.650 & 22.539 & -23.383  & \citet{Ike11} \\
1163086 & 150.703770 & 2.370019 & 22.444 & 3.748 & 22.863 & -23.122  & \citet{Mar16a} \\
1208399 & 150.259540 & 2.376141 & 21.424 & 3.717 & 21.488 & -24.478  & \citet{Mar16a} \\
1224733 & 150.208990 & 2.438466 & 21.147 & 3.715 & 21.485 & -24.480  & \citet{Mar16a} \\
1273346\tablenotemark{*} & 149.776910 & 2.444306 & 22.779 & 4.170 & 23.274 & -22.952  & \citet{Mar16a} \\
1730531\tablenotemark{*} & 149.843220 & 2.659095 & 22.900 & 3.748 & 23.439 & -22.545  & our spectroscopy \\
1856470 & 150.475680 & 2.798362 & 21.282 & 4.110 & 21.753 & -24.445  & \citet{Mar16a} \\
1938843 & 149.845860 & 2.860459 & 22.160 & 3.630 & 22.619 & -23.290  & our spectroscopy \\
1971812 & 149.472870 & 2.793400 & 21.887 & 3.610 & 22.179 & -23.717  & \citet{Mar16a} \\
\enddata
\tablenotetext{*}{Not included in the space density bins because M$_{1450}>$-23.0}
\end{deluxetable*}

\section{Space Density Determination}

The advantage of doing this study in the COSMOS field is that it already contains extensive 
spectroscopic follow-up and extensive multi-wavelength data from radio to X-rays. Thus, 
combining the confirmed candidates presented above, with known AGNs from the literature, 
we obtain a sample of 16 spectroscopically confirmed AGNs with 3.6$<z<$4.2 and i$_{AB}<$23, 
presented in Table \ref{tab:spaced}, along with the relative r$_{AB}$ for each source. The 
i$_{AB}$ magnitudes are from HST F814W band, while the r$_{AB}$ magnitudes are from the 
Subaru telescope and they are described in \citet{Ilb09}.

The absolute magnitude at 1450 {\AA} rest frame (M$_{1450}$) for each source has been 
derived from the $r_{AB}$ magnitude applying a K-correction according to the following 
formula:
\begin{equation}
M_{1450} = r_{AB} - 2.5log(1+z_{spec})+K_{corr} 
\end{equation}
where
\begin{equation}
K_{corr}=2.5\alpha_\nu log_{10}(\lambda_{obs}/(1+z_{spec})/\lambda_{rest})
\end{equation}
The AGN intrinsic slope $\alpha_\nu$ is fixed to -0.7, while $\lambda_{obs}=6284$ {\AA} 
is the central wavelength of the $r_{AB}$ filter and $\lambda_{rest}=1450$ {\AA}. The reason 
we chose the $r_{AB}$ band is because it is not affected by strong quasar emission lines, 
like C$_{IV}$ that falls in the $i_{AB}$ band for this redshift range. The K-correction 
is redshift dependent and ranges from 0.05mag at z=3.6 to 0.14mag at z=4.2. To check the 
robustness of our absolute magnitude determination, we have used various methods to 
calculate it (using the i and r band, with and without K-correction). The point at 
M$_{1450}$=-24.5 remains basically unaltered, and we only see small changes in it at 
M$_{1450}$=-23.5, with the current determination resulting in the faintest absolute 
magnitudes, thus being the most conservative.

We find four sources for -25$<M_{1450}<$-24 and 9 for -24$<M_{1450}<$-23. Based on these 
numbers, we calculate the space density of AGNs at this redshift range in the two magnitude 
bins. This space density, summarized in Table \ref{tab:spacedn}, has been derived by 
dividing the actual number of the spectroscopically confirmed AGNs with the comoving 
volume between $3.6\leq z\leq 4.2$, without any correction for incompleteness. Thus it 
represents a robust lower limit to the real space density of $z\sim 4$ AGNs. As can be 
seen in Figure 3, without any corrections (blue filled squares), our measurements agree 
well with the \citet{Gli11} analysis and put more stringent constraints on the knee 
of the LF.

Considering completeness and contamination, we can estimate rough corrections. 
In the color selection, we have six AGNs with 3.6$<z_{spec}<$4.2 out of 12 AGNs with 
known $z_{spec}$. Thus, out of the 11 candidates without $z_{spec}$, we expect five or six 
($\sim$50\%) to have 3.6$<z_{spec}<$4.2, resulting to 11 or 12 potential AGNs with the 
(B$_{J}$-V$_{J}$) versus (r-i) selection. The known AGNs with 3.6$<z_{spec}<$4.2 and 
i$_{AB}\leq$23.0 are 16, of which we recover 37.5\% with the (B$_{J}$-V$_{J}$) versus 
(r-i) criterion (six out of 16). This means that the total number of AGNs expected in 
COSMOS at 3.6$<z_{spec}<$4.2 and i$_{AB}<$23.0 could be N$_{tot}$=11$\times$16/6=29.3 or 
N$_{tot}$=12$\times$16/6=32. The space density determination using only the 16 AGNs known 
at 3.6$<z_{spec}<$4.2, i$_{AB}\leq$23.0 and -26.0$<M_{1450}<$-23.0 is incomplete by a factor 
of 0.50-0.55 at least. If we correct for this incompleteness factor, we will go to a level 
higher than \cite{Gli11} (green open squares in Figure 3). Adopting a slightly different 
color criterion, with ($B_{J}-V_{J})>1.1$ instead of 1.3 as threshold, the expected total 
number of AGNs is 34 and the completeness corrections remain at the $\sim 50\%$ level. 
This indicates that the green squares (corrected space density) in Figure 3 are quite 
robust with respect to the details of the adopted color criterion.

\begin{deluxetable*}{cccccc}
\tablecaption{AGN Space Density ($<z>$=3.9)\label{tab:spacedn}}
\tablehead{
\colhead{M$_{1450}$} & \colhead{$\Phi$} &
\colhead{$\sigma_\Phi^{up}$} & \colhead{$\sigma_\Phi^{low}$} & 
\colhead{N$_{AGN}$} & \colhead{$\Phi_{corr}$}\\ & $Mpc^{-3} Mag^{-1}$ & & & & \\}
\startdata
-24.5 & 3.509e-07& 2.789e-07& 1.699e-07& 4 &7.018e-07\\
-23.5 &7.895e-07& 3.616e-07& 2.595e-07& 9 & 1.579e-06 \\
\enddata
\end{deluxetable*}

In Figure 3 we also present the LFs calculated by \citet{Aki18}, \citet{Par18}, and 
\citet{Mas12} for comparison. The sample created by \citet{Aki18} is limited to g-band 
dropout (i.e. $3.5<z<4.0$) point-like sources, for which they have derived photometric 
redshifts based on five filters (g,r,i,z,y). The faint-end slope presented in 
their work is too shallow to be reconciled with our measurements, which correspond to 
spectroscopically confirmed sources. The LF by \citet{Ike11} is also lower than our space 
density determination. On the other hand \citet{Par18}, after performing an independent 
photometric redshift estimate of the X-ray-selected sample presented by \citet{Gia15}, 
discarded 10 of the 22 sources that were supposed to lie at $z>$4. Even though the faint-end 
slope in \citet{Par18} is steeper than that found by \citet{Gli11}, their space density in 
absolute magnitudes M$_{1450}<$-23 is marginally in agreement with our estimates. We also 
show the space density derived by \citet{Mar16b}, based on X-ray data, after being converted 
to UV \citep{Ric17}. Although these points are higher than most optical LFs, they are slightly 
lower than our estimate. In Table \ref{tab:spacedn} we present the estimate of the AGN space 
density $\Phi$, based on our analysis, in the two absolute magnitude bins. Even excluding 
the COSMOS247934, which is the least certain among our sources, the uncorrected space density 
at M$_{1450}$=-23.5 becomes 7.018e-07 $Mpc^{-3}Mag^{-1}$, which is still higher than all LFs 
presented in Figure 3, except for \citet{Par18}. Considering the space density corrected for
incompleteness, also the \citet{Par18} LF also turns out to be underestimated.

An important aspect, made clear by our sample, is that selections based on color criteria 
can be highly incomplete, since out of the 16 spectroscopically confirmed AGNs only six 
have been selected by color. So far, the majority of studies on the AGN LF at this redshift 
range is based on color-selected samples and this could be the reason why faint AGN number 
densities have been underestimated. When a first attempt was made by \citet{Gia15} to create 
an AGN sample based on non-traditional criteria, a different picture emerged. Given that AGNs, 
even at faint magnitudes, have a large escape fraction as shown by \citet{Gra18}, an increase 
of the estimate of their population can have significant implications on the contribution 
of AGNs to the H$_{I}$ ionizing background.  

\section{Discussion and Conclusions}

Our estimates of the space density in the range $-24.5<M_{1450}<-23.5$, shown by filled 
blue squares in Figure 3, are not corrected for any incompleteness factor, thus they 
represent firm lower limits, assuming that the density fluctuations due to cosmic variance 
are not important.

To explore this possibility, we checked how the volume density of AGNs at $z\sim$4 in COSMOS 
compares to the average derived from the SDSS survey. In the COSMOS area there is no known 
quasi stellar object (QSO) brighter than i$_{AB}$=21.0 in the redshift interval $3.6<z<4.2$. 
Considering the SDSS DR14 catalog \citep{Pas18} in areas of different sizes, ranging from 
10-100 deg$^{2}$, centered on the COSMOS field, we find a mean density of $0.59\pm 0.10~deg^{-2}$ 
compared to the mean SDSS density of $0.61\pm 0.01~deg^{-2}$ (5683 QSOs in 9376 deg$^{2}$).
This indicates that the COSMOS field is not particularly overdense or underdense at $z\sim 4$. 
For this reason, the completion of a spectroscopic AGN survey in this field, such as the 
one presented here, is fundamental to address the role of AGNs in the reionization epoch. 

Indeed, the AGN space density derived by our study is a conservative estimate and might 
increase in the future. As can be seen in Table \ref{tab:colorsel}, six confirmed AGNs at 
$z_{spec}\geq 3$ have $z_{phot}<$1 in \cite{Ilb09}. In two cases, where both the spectroscopic 
and the photometric redshifts are larger than 3, the $z_{phot}$ tends to systematically 
underestimate the actual $z_{spec}$. This is an indication that our selection based on 
$z_{phot}$ could still be biased against $z\sim 4$ AGNs, and that different $z_{phot}$ recipes 
could increase the number of AGN candidates selected by photometric redshifts. As an example, 
\citet{Sal11} provided photometric redshifts only for two of our sources in 
Table \ref{tab:colorsel}, i.e., source id=330806 with $z_{phot}=3.949$ and id=1226535 
with $z_{phot}=4.545$. Their estimates are in reasonable agreement with the spectroscopic 
redshifts of these AGNs.

\begin{figure*}
\label{fig:spDen}
\plotone{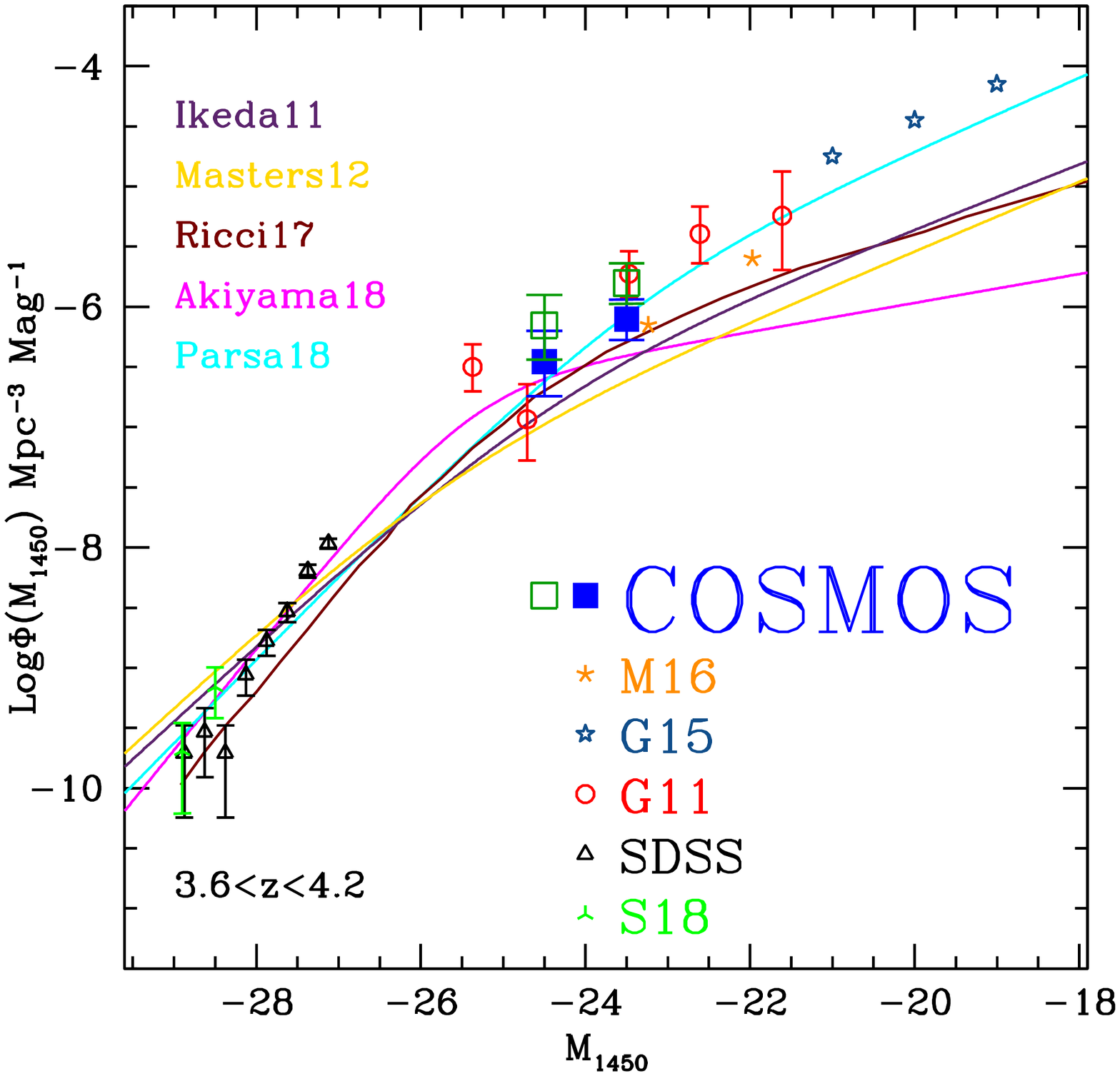}
\caption{Luminosity function of quasi stellar objects/AGNs at $z\sim$4. Black triangles 
show the bright end determined by the SDSS \citep{Aki18}, green triangles are the points 
presented by \citet{Sch18}, red circles show the bins calculated by \citet{Gli11}, 
blue stars show the bins by \citet{Gia15}, filled blue squares represent the two magnitude 
bins from this work (no corrections), and the open green squares correspond to the space 
density derived in this work after applying the corrections discussed in Section 4. Notice 
that our data actually compensate for the apparent dip present in the work by \citet{Gli11}. 
The results by \citet{Mar16b}, converted into UV, are presented as orange asterisks. For 
comparison we also present the LFs derived by \citet{Ike11}, \citet{Mas12}, \citet{Ric17}, 
\citet{Aki18}, and \citet{Par18}. All LFs have been evolved to z=3.9 following the density 
evolution law by \citet{Fan01}.}
\end{figure*}

\citet{Stev18}, in their analysis of the UV LF of a mixed sample including both SFGs and 
galaxies dominated by AGNs at $z\sim$4 in the SHELA survey, found a space density of 
$10^{-6} Mpc^{-3}$ at $M_{1450}=-23.5$. Our measurements in the same redshift and absolute 
magnitude is $\sim1.6\times10^{-6} Mpc^{-3}$ (corrected), and this is an indication that 
their LF is dominated by AGNs at $M_{1450}<-23.5$. Based on their discussion, a high AGN 
space density would mean that AGNs could be largely responsible for the H$_{I}$ ionizing 
UVB at $z=4$.

We are confident that on the net of all random and systematic effects, the corrected 
estimate of the LF presented in this work represents a robust determination of the space 
density of L$^{*}$ AGNs at $z\sim 4$. The agreement of our measurement with the \citet{Gli11} 
results favors at $M_{1450}<-23.5$ a steeper slope of the LF for the COSMOS field than that 
determined by \citet{Ike11}, \citet{Mas12}, \citet{Ric17}, \citet{Aki18}, and \citet{Par18}. 

This study poses an open challenge that should be addressed in the future with major 
observational effort: to measure with great accuracy the space density of AGNs at L=L$^{*}$ 
and $z>4$. In fact, in a subsequent work, once our spectroscopic sample is complete, we will 
present the global shape of the LF at $z\sim$4 and the associated emissivity. This can have 
deep implications on the extrapolation of the number of QSOs expected at high-z in wide and 
deep large area surveys, either ground based, e.g. LSST, or from space, e.g., e-Rosita, 
Euclid, WFIRST. An upward revision of the number density of L=L$^{*}$ AGNs would certainly 
imply a reconsideration of the expected QSO and AGN numbers at $z>4$ in these future missions. 

\acknowledgments
We would like to thank the anonymous referee for useful suggestions and constructive 
comments that helped us improve this paper. This paper includes data gathered with the 
6.5 meter Magellan Telescopes located at Las Campanas Observatory (LCO), Chile.


\begin{thebibliography}{}

\bibitem[Akiyama et al.(2018)]{Aki18} Akiyama, M., He, W., Ikeda, H., et al.\ 2018, \pasj, 70, S34 
\bibitem[Becker \& Bolton(2013)]{BeB13} Becker, G.~D., \& Bolton, J.~S.\ 2013, \mnras, 436, 1023 
\bibitem[Becker et al.(2015)]{Bec15} Becker, G.~D., Bolton, J.~S., Madau, P., et al.\ 2015, \mnras, 447, 3402
\bibitem[Becker et al.(2018)]{Bec18} Becker, G.~D., Davies, F. B., Furlanetto, S. R., et al.\ 2018, arXiv:1803.08932
\bibitem[Bian et al.(2017)]{Bian17} Bian, F., Fan, X., McGreer, I., Cai, Z., \& Jiang, L.\ 2017, \apjl, 837, L12 
\bibitem[Bongiorno et al.(2007)]{Bon07} Bongiorno, A., Zamorani, G., Gavignaud, I., et al.\ 2007, \aap, 472, 443 
\bibitem[Bosman et al.(2018)]{Bos18} Bosman, S.~E.~I., Fan, X., Jiang, L., et al.\ 2018, \mnras, 479, 1055 
\bibitem[Bouwens et al.(2016)]{Bow16} Bouwens, R.~J., Smit, R., Labb{\'e}, I., et al.\ 2016, \apj, 831, 176 
\bibitem[Brusa et al.(2009)]{Bru09} Brusa, M., Comastri, A., Gilli, R., et al.\ 2009, \apj, 693, 8 
\bibitem[Chardin et al.(2015)]{Cha15} Chardin, J., Haehnelt, M.~G., Aubert, D., \& Puchwein, E.\ 2015, \mnras, 453, 2943 
\bibitem[Chardin et al.(2017)]{Cha17} Chardin, J., Puchwein, E., \& Haehnelt, M.~G.\ 2017, \mnras, 465, 3429 
\bibitem[Civano et al.(2012)]{Civ12} Civano, F., Elvis, M., Brusa, M., et al.\ 2012, \apjs, 201, 30 
\bibitem[Civano et al.(2016)]{Civ16} Civano, F., Marchesi, S., Comastri, A., et al.\ 2016, \apj, 819, 62 
\bibitem[Cowie et al.(2009)]{Cow09} Cowie, L.~L., Barger, A.~J., \& Trouille, L.\ 2009, \apj, 692, 1476 
\bibitem[Cristiani et al.(2016)]{Cri16} Cristiani, S., Serrano, L.~M., Fontanot, F., Vanzella, E., \& Monaco, P.\ 2016, \mnras, 462, 2478 
\bibitem[Croom et al.(2009)]{Cro09} Croom, S.~M., Richards, G.~T., Shanks, T., et al.\ 2009, \mnras, 399, 1755 
\bibitem[D'Aloisio et al.(2017)]{DAl17} D'Aloisio, A., Upton Sanderbeck, P.~R., McQuinn, M., Trac, H., \& Shapiro, P.~R.\ 2017, \mnras, 468, 4691
\bibitem[Dressler et al.(2011)]{Dre11} Dressler, A., Bigelow, B., Hare, T., et al.\ 2011, \pasp, 123, 288 
\bibitem[Fan et al.(2001)]{Fan01} Fan, X., Strauss, M.~A., Schneider, D.~P., et al.\ 2001, \aj, 121, 54 
\bibitem[Fan et al.(2006)]{Fan06} Fan, X., Strauss, M.~A., Becker, R.~H., et al.\ 2006, \aj, 132, 117 
\bibitem[Finkelstein et al.(2015)]{Fin15} Finkelstein, S.~L., Ryan, R.~E., Jr., Papovich, C., et al.\ 2015, \apj, 810, 71 
\bibitem[Fletcher et al.(2018)]{Fle18} Fletcher, T.~J., Robertson, B.~E., Nakajima, K., et al.\ 2018, arXiv:1806.01741 
\bibitem[Giallongo et al.(2015)]{Gia15} Giallongo, E., Grazian, A., Fiore, F., et al.\ 2015, \aap, 578, A83 
\bibitem[Glikman et al.(2011)]{Gli11} Glikman, E., Djorgovski, S.~G., Stern, D., et al.\ 2011, \apjl, 728, L26 
\bibitem[Grazian et al.(2018)]{Gra18} Grazian, A., Giallongo, E., Boutsia, K., et al.\ 2018, \aap, 613, A44 
\bibitem[Haardt \& Madau(2012)]{HM12} Haardt, F., \& Madau, P.\ 2012, \apj, 746, 125 
\bibitem[Hasinger et al.(2018)]{Hasin18} Hasinger, G., Capak, P., Salvato, M., et al.\ 2018, \apj, 858, 77 
\bibitem[Hassan et al.(2018)]{Has18} Hassan, S., Dav{\'e}, R., Mitra, S., et al.\ 2018, \mnras, 473, 227 
\bibitem[Ikeda et al.(2011)]{Ike11} Ikeda, H., Nagao, T., Matsuoka, K., et al.\ 2011, \apjl, 728, L25 
\bibitem[Ilbert et al.(2009)]{Ilb09} Ilbert, O., Capak, P., Salvato, M., et al.\ 2009, \apj, 690, 1236 
\bibitem[Jones et al.(2018)]{Jon18} Jones, L.~H., Barger, A.~J., Cowie, L.~L., et al.\ 2018, \apj, 862, 142 
\bibitem[Keating et al.(2018)]{Kea18} Keating, L.~C., Puchwein, E., \& Haehnelt, M.~G.\ 2018, \mnras, 477, 5501 
\bibitem[Khaire et al.(2016)]{Kha16} Khaire, V., Srianand, R., Choudhury, T.~R., \& Gaikwad, P.\ 2016, \mnras, 457, 4051 
\bibitem[Madau \& Haardt(2015)]{MH15} Madau, P., \& Haardt, F.\ 2015, \apjl, 813, L8 
\bibitem[Marchesi et al.(2016a)]{Mar16a} Marchesi, S., Civano, F., Elvis, M., et al. 2016a, ApJ 817 34 
\bibitem[Marchesi et al.(2016b)]{Mar16b} Marchesi, S., Civano, F., Salvato, M., et al.\ 2016b, \apj, 827, 150 
\bibitem[Masters et al.(2012)]{Mas12} Masters, D., Capak, P., Salvato, M., et al.\ 2012, \apj, 755, 169 
\bibitem[Naidu et al.(2018)]{Nai18} Naidu, R.~P., Forrest, B., Oesch, P.~A., Tran, K.-V.~H., \& Holden, B.~P.\ 2018, \mnras, 478, 791 
\bibitem[P{\^a}ris et al.(2018)]{Pas18} P{\^a}ris, I., Petitjean, P., Aubourg, {\'E}., et al.\ 2018, \aap, 613, A51 
\bibitem[Parsa et al.(2018)]{Par18} Parsa, S., Dunlop, J.~S., \& McLure, R.~J.\ 2018, \mnras, 474, 2904 
\bibitem[Planck Collaboration et al.(2018)]{Planck} Planck Collaboration, Akrami, Y., Arroja, F., et al.\ 2018, arXiv:1807.06205 
\bibitem[Puchwein et al.(2018)]{Puc18} Puchwein, E., Haardt, F., Haehnelt, M.~G., \& Madau, P.\ 2018, arXiv:1801.04931 
\bibitem[Ricci et al.(2017)]{Ric17} Ricci, F., Marchesi, S., Shankar, F., La Franca, F., \& Civano, F.\ 2017, \mnras, 465, 1915 
\bibitem[Salvato et al.(2011)]{Sal11} Salvato, M., Ilbert, O., Hasinger, G., et al.\ 2011, \apj, 742, 61 
\bibitem[Schindler et al.(2018)]{Sch18} Schindler, J.-T., Fan, X., McGreer, I.~D., et al.\ 2018, \apj, 863, 144 
\bibitem[Shapley et al.(2016)]{Sha16} Shapley, A.~E., Steidel, C.~C., Strom, A.~L., et al.\ 2016, \apjl, 826, L24
\bibitem[Steidel et al.(2002)]{Ste02} Steidel, C.~C., Hunt, M.~P., Shapley, A.~E., et al.\ 2002, \apj, 576, 653 
\bibitem[Steidel et al.(2018)]{Ste18} Steidel, C.~C., Bogosavlevic, M., Shapley, A.~E., et al.\ 2018, arXiv:1805.06071 
\bibitem[Stevans et al.(2018)]{Stev18} Stevans, M.~L., Finkelstein, S.~L., Wold, I., et al.\ 2018, \apj, 863, 63 
\bibitem[Tanvir et al.(2018)]{Tan18} Tanvir, N.~R., Fynbo, J.~P.~U., de Ugarte Postigo, A., et al.\ 2018, arXiv:1805.07318 
\bibitem[Trump et al.(2009)]{Tru09} Trump, J.~R., Impey, C.~D., Elvis, M., et al.\ 2009, \apj, 696, 1195 
\bibitem[Vanzella et al.(2016)]{Van16} Vanzella, E., de Barros, S., Vasei, K., et al.\ 2016, \apj, 825, 41 

\end{thebibliography}
\end{document}